\documentclass[%
 reprint,
 amsmath,amssymb,
 aps,
]{revtex4-2}

\usepackage{graphicx}
\usepackage{dcolumn}
\usepackage{bm}
\usepackage{wrapfig}
\usepackage{comment}
\usepackage{ esint }
\usepackage{soul}
\usepackage[unicode]{hyperref}
\usepackage[dvipsnames]{xcolor}
\usepackage{color,soul}

\begin{document}

\title{Superfluid transition of disordered dipolar Fermi gases in a 2D lattice}

\author{V. Y. Pinchenkova,$^{1,2}$ S. I. Matveenko,$^{3,1}$ and G. V. Shlyapnikov$^{1,2,4,5}$}
    \affiliation{$^{1}$Russian Quantum Center, Skolkovo, Moscow 143025, Russia}
    \affiliation{$^{2}$Moscow Institute of Physics and Technology, Dolgoprudny, Moscow Region, 141701, Russia}
    \affiliation{$^{3}$L. D. Landau Institute for Theoretical Physics, Chernogolovka, Moscow region 142432, Russia}

    \affiliation{$^{4}$Université Paris-Saclay, CNRS, LPTMS, 91405 Orsay, France}
    \affiliation{$^{5}$Van der Waals–Zeeman Institute, Institute of Physics, University of Amsterdam, Science Park 904, 1098 XH Amsterdam, The Netherlands}
\date{\today}

\begin{abstract}

We consider a superfluid transition in two-component dipolar Fermi gases in a two-dimensional lattice with a weak on-site disorder. The momentum dependent dipole-dipole interaction amplitude violates the Anderson theorem and in the weakly interacting regime this leads to an increase of the superfluid transition temperature. We find that in a sufficiently deep lattice (tight-binding regime) and in the low momentum limit superfluid properties can be considered in the same way as in free space replacing the mass of atoms by an effective mass in the lattice. The disorder-induced increase of the critical temperature can be significantly more pronounced than in free space.

\end{abstract}

\maketitle
\section{Introduction}\label{sec1}

The success in studying atomic Fermi gases in the last decades relies on the use of Feshbach resonances (magnetic field dependence of the interaction strength), which allows one to change the interaction between atoms in a wide range \cite{feshb}. Experiments with two-component Fermi-gases have obtained the strongly interacting regime and a superfluid transition in this regime \cite{supertrans1,supertrans2}. However, the weakly interacting Bardeen-Cooper-Schrieffer (BCS) regime has not been achieved because at realistic densities the required critical temperature is about a nanokelvin or lower, which is beyond experimental reach. Possibilities to manipulate the superfluid transition temperature rely on varying the external confining potential \cite{supertrans2}, in particular by
introducing a disorder \cite{Sanchez-Palencia,ReviewHuse,ReviewAbanin,anderson,AG, aa,lee,sad,ov,fin,temp-increase,T&K,Stri,Dis-at-AT,disorder,Roati}. However, in a weak disorder ($k_F l \gg 1$, where $k_F$ is the Fermi momentum, and $l$ the mean free path) in the case of a weak short-range interaction where the interaction amplitude is momentum independent, the Anderson theorem indicates that the  BCS transition temperature does not depend on the disorder \cite{anderson,AG}. On the other hand, weak localization effects \cite{aa,lee,fin}, which in the presence of interaction change the fermion self-energy and the density of states,  lead to a small increase of the BCS transition temperature in the absence of  Coulomb interaction \cite{temp-increase,T&K}.

The situation drastically changes for Fermi gases of dipolar particles, where the interaction amplitude is momentum dependent and the Anderson theorem is violated \cite{Superfluid}. In two-dimensional (2D) two-component fermions with dipoles perpendicular to the plane of their translational motion (magnetic atoms or polar molecules) the interaction amplitude consists of a fairly large repulsive contribution and a long-range attractive momentum-dependent contribution. However, the short-range repulsion can be strongly reduced or even converted to attraction by using Feshbach resonances, so that the resulting amplitude becomes attractive. This provides a superfluid transition. The transition temperature in the weakly interacting regime can be strongly increased by a weak disorder (by a factor of 2 or more for realistic parameters \cite{Superfluid}), although an increase in the interaction strength decreases the influence of disorder and in the strongly interacting regime this influence practically vanishes \cite{our}.

In the present paper we consider two-component fermions in the 2D lattice with a weak on-site disorder, assuming that the dipole moments of fermions are perpendicular to the lattice. In a sufficiently deep lattice and in the low momentum limit, the superfluid pairing of fermions is similar to that in free space, but replacing the atom mass $m$ with an effective mass $m_*$. In the weakly interacting regime the presence of disorder strongly increases the critical temperature, and this effect can be more pronounced than without a lattice.

This paper is organized as follows. In Sec.\ref{sec2} we introduce the renormalized BCS gap equation in a 2D lattice, which takes into account a weak on-site disorder. Sec.\ref{sec3} is devoted to the s-wave scattering amplitude of dipolar fermions. In Sec.\ref{sec4} we obtain numerical solutions of the gap equation in the absence and presence of the weak disorder for different values of the effective mass and compare the result in the lattice with the result in free space. We conclude in Sec.\ref{sec5}.

\section{Gap equation}\label{sec2}
We consider a two-component dipolar Fermi gas confined within a 2D periodic lattice potential $U(\mathbf{r})$. We begin with presenting relations in the absence of disorder. In terms of the field operators of fermionic components $\hat{\Psi}_{\uparrow} (\mathbf{r})$ and $\hat{\Psi}_{\downarrow} (\mathbf{r})$ the Hamiltonian is given by

\begin{equation}\label{eq1}
\begin{gathered}
\hat{H} = \int d^2 \mathbf{r} \sum_{\alpha = \uparrow, \downarrow} \hat{\Psi}^{\dagger}_\alpha(\mathbf{r}) \left ( -\frac{\nabla^2}{2 m}  - \mu +U(\mathbf{r}) \right )\hat{\Psi}_\alpha(\mathbf{r}) + \\ \frac{1}{2} \int d^2 \mathbf{r} \: d^2 \mathbf{r'}  \sum_{\alpha, \beta = \uparrow, \downarrow}
\hat{\Psi}^{\dagger}_{\alpha} (\mathbf{r}) \hat{\Psi}^{\dagger}_{\beta} (\mathbf{r'}) V(\mathbf{r -r'})\hat{\Psi}_{\beta} (\mathbf{r'}) \hat{\Psi}_{\alpha} (\mathbf{r}),
\end{gathered}
\end{equation}
where $m$ is the particle mass, $\mu$ is the chemical potential, $V(\mathbf{r -r'})$ is an effective attractive interaction potential, and we put $\hbar = 1$.

In the 2D geometry, the temperature for the onset of superfluidity $T_c$ is set by the Berezinskii-Kosterlitz-Thouless (BKT) mechanism. In the weakly interacting regime, however, the BKT transition temperature is very close to $T_c$ calculated in the mean-field BCS approach for both, gases without disorder \cite{Miyake} and gases with a weak disorder \cite{our}. To find the BCS gap equation that determines $T_c$, we first write the field operators in the Bloch basis:

\begin{equation}\label{eq2}
\hat{\Psi}_{\alpha} (\mathbf{r}) = \sum_{\nu, \mathbf{k}} \hat{a}_{\nu \mathbf{k}\alpha} \chi_{\nu \mathbf{k}}(\mathbf{r}),
\end{equation}
where the Bloch functions $\chi_{\nu \mathbf{k}}(\mathbf{r})$ are determined by the Schr$\ddot{\textup{o}}$dinger equation:

\begin{equation}\label{eq3}
\left[ -\frac{\nabla^2}{2 m} +U(\mathbf{r})\right] \chi_{\mathbf{k} \nu}(\mathbf{r}) = E_{\mathbf{k} \nu} \chi_{\mathbf{k} \nu}(\mathbf{r}),
\end{equation}
with $E_{\mathbf{k} \nu}$  being a single particle energy in an energy band numerated by $\nu = 0, 1, 2...$, $\hat{a}_{\nu \mathbf{k}\alpha}$, and $\hat{a}^{\dagger}_{\nu \mathbf{k}\alpha}$ the annihilation and creation operators  of fermions with
quasimomentum $\mathbf{k} = \{k_x, k_y \}$ taking values within the Brillouin
zone: $\{-\pi/b<k_i<\pi/b; \:i = x, y\}$, and $b$ the lattice period.

We consider a sufficiently deep lattice and ultralow temperatures, which leads to the fact that all fermions are in the lowest energy band $\nu = 0$. We also consider a small filling factor (low momentum limit), which means that the number of particles is much smaller than the number of lattice sites. This requirement results in the condition $k_F b \ll 1$ with $k_F = \sqrt{2 \pi n}$ being the Fermi momentum, and $n$ the particle density. Thus, in the low momentum limit the Fermi energy $E_F$ is much smaller than the energy bandwidth, and hence the single particle dispersion relation can be approximated by

\begin{equation}\label{eq4}
\begin{gathered}
E_k = \frac{k^2}{2 m_*},
\end{gathered}
\end{equation}
with $m_*$ being an effective mass. Hereinafter we omit the corresponding index $\nu = 0$.

We rewrite $\hat{H}$ (\ref{eq1}) in the described Bloch basis and reduce it to a standard BSC form:

\begin{equation}\label{eq5}
\begin{gathered}
\hat{H} = \sum_{\mathbf{k}, \: \alpha = \uparrow, \downarrow} \xi_k \: \hat{a}^{\dagger}_{\mathbf{k}\alpha} \hat{a}_{\mathbf{k}\alpha} + \sum_{\mathbf{k}} \left[ \hat{a}^{\dagger}_{\mathbf{k}\uparrow}  \: \hat{a}^{\dagger}_{-\mathbf{k} \downarrow} \Delta(\mathbf{k}) + \text{h.c.} \right ],
\end{gathered}
\end{equation}
where $\xi_k = E_k - \mu$, and in the weakly interacting regime $\mu \approx E_F$ with $E_F = k_F^2/(2m_*)$. We neglected the intra-species interaction in Eq. (\ref{eq5}) because it renormalizes the chemical potential, which does not influence the critical temperature $T_c$. Second, the intra-species interaction introduces a fairly small change of the effective mass,
which only slightly depends on the disorder, and the influence of the
disorder-dependent part of the effective mass on $T_c$ can be omitted. The superfluid order parameter (gap) $\Delta(\mathbf{k})$ obeys the gap equation

\begin{equation}\label{eq6}
\Delta(\mathbf{k}) = \sum_{\mathbf{k'}} V(\mathbf{k}, \mathbf{k}') \left< \hat{a}_{\mathbf{k}'\downarrow} \hat{a}_{\mathbf{-k}'\uparrow} \right>,
\end{equation}
where the symbol $\left< ...\right>$ denotes the statistical average, and $V(\mathbf{k}, \mathbf{k}')$ is the matrix element of the interaction
potential:

\begin{equation}\label{eq7}
V(\mathbf{k}, \mathbf{k}') = \int d^2 \mathbf{r} \: d^2 \mathbf{r'} \chi_{\mathbf{k}}^{*}(\mathbf{r}) \chi_{-\mathbf{k}}^{*}(\mathbf{r}') V(\mathbf{r} - \mathbf{r}') \chi_{\mathbf{k}'}(\mathbf{r}') \chi_{-\mathbf{k}'}(\mathbf{r}).
\end{equation}

We now rewrite the gap equation (\ref{eq6}) in terms of the finite-temperature Green functions:

\begin{equation}\label{eq8}
\Delta(\mathbf{k}) =  -T \sum_{\omega_j} \int \frac{d^2 \mathbf{k'}}{(2 \pi)^2} V(\mathbf{k}, \mathbf{k}') F (\mathbf{k'}, \omega_j),
\end{equation}
where the normal and anomalous Green functions in the Heisenberg representation with an imaginary time $\tau$ are given by $G(\mathbf{k}, \tau) =  -\left<T_\tau \hat{a}_{\mathbf{k}\alpha} (\tau) \: \hat{a}^{\dagger}_{\mathbf{k}\alpha} (0) \right>$ and $F(\mathbf{k}, \tau) =  -\left<T_\tau \hat{a}_{\mathbf{k}\uparrow} (\tau) \: \hat{a}_{-\mathbf{k}\downarrow} (0) \right>$, and we have used the Fourier transform $F(\mathbf{k}, \tau) = T \sum_{\omega_j} F (\mathbf{k'}, \omega_j) e^{-i \omega_j \tau}$. The summation runs over the fermion Matsubara frequencies $\omega_j = \pi T(2j+1)$, $j = 0, \pm 1, ...$.

In the next step we include a weak disorder of the form

\begin{equation}\label{eq9}
\hat{H}_{dis} =  \int d^2 \mathbf{r} \sum_{\alpha = \uparrow, \downarrow} \hat{\Psi}^{\dagger}_\alpha(\mathbf{r}) U_{dis}(\mathbf{r})  \hat{\Psi}_\alpha(\mathbf{r}),
\end{equation}
with the short-range on-site disorder potential:

\begin{equation}\label{eq10}
U_{dis}(\mathbf{r}) = \sum_i \eta_i u_0 \:\delta(\mathbf{r} - \mathbf{R}_i),
\end{equation}
where $\mathbf{R}_i$ are the coordinates of the lattice sites, and the random variable $\eta$ takes values from -1 to 1 with a probability $P(\eta) = \theta(1 - |\eta|)/2$. The variable $\eta$ has a vanishing average:

\begin{equation}\label{eq11}
\left <\eta \right>_{dis} = \int P(\eta) \eta \: d\eta = 0,
\end{equation}
and the variance $\left<\eta_i \eta_j\right>_{dis} = \delta_{ij}/3$.

To treat this weak disorder we make use of the mean-field Abrikosov and Gor’kov approach \cite{AG}. This method involves a modification of the gap equation (\ref{eq8}) through the substitution of ordinary Green functions with the Green functions $\left < G (\mathbf{k}, \omega_j)\right>_{dis} \equiv \bar{G}(\mathbf{k}, \omega_j)$ and $\left < F (\mathbf{k}, \omega_j) \right>_{dis} \equiv \bar{F}(\mathbf{k}, \omega_j)$ averaged over the disorder and given in Appendix by Eqs. (\ref{eq12}), (\ref{eq13}). The solutions for these functions are obtained in Appendix (Eqs. (\ref{eq16})- (\ref{eq19})).

We then use the relation between the fermion-fermion scattering amplitude $f(\mathbf{k'}, \mathbf{k})$ and the matrix element $V(\mathbf{k}, \mathbf{k}')$ and find the renormalized gap equation averaged over the disorder (see Ref. \cite{our} and references therein):

\begin{equation}\label{eq20}
\Delta(k) =  - \fint \frac{k' dk'}{2 \pi} f  (k', k) \Delta(k')\left [K(k') - \frac{1}{2 (E_{k'} - E_k)} \right ],
\end{equation}
where $f(\mathbf{k'}, \mathbf{k})$ and $\Delta(\mathbf{k})$ are replaced by their s-wave parts $f(k',k)$ and $\Delta(k)$. The main contribution to the integral over $dk'$ comes from the region near the Fermi surface, we thus can put $\Delta(k)=\Delta(k_F)$ under the integral. In the BCS limit $\Delta(k_F) \rightarrow 0$ and for $k = k_F$, the gap equation becomes

\begin{equation}\label{eq21}
1 =  -\fint \frac{k 'd k'}{2 \pi} f  (k',k_F) \left [K(k') - \frac{1}{2 (E_{k'} - E_F)} \right ]
\end{equation}
where the function $K(k)$ reads

\begin{equation}\label{eq22}
\begin{split}
K(k) & = 2 T_c \sum_{\omega_j>0}\frac{1 + \frac{1}{2 \tau_e \omega_j}}{  \xi^2_{k} + (\omega_j + \frac{1}{2 \tau_e})^2} \\ & = \frac{i}{2 \pi} \frac{ \Psi( \frac{1}{2} - \frac{i z_k}{2 \pi T_c}) - \Psi(\frac{1}{2})}{z_k} + \text{h.c.},
\end{split}
\end{equation}
where $\Psi$ is the digamma function, and $z_k = \xi_k + i/(2 \tau_e)$. The quantity $\tau_e$ is the inverse disorder-induced elastic scattering rate $1/\tau_e = 2 \pi \nu \gamma$ with $\nu = m_*/2\pi$ being the 2D density of states, and the factor $\gamma$ comes from the correlation function for the disorder potential $\left < U_{dis}(\mathbf{r}) U_{dis}(\mathbf{r'})\right >_{dis} = \gamma \delta (\mathbf{r}-\mathbf{r'})$. The weak disorder means that $E_F \tau_e \gg 1$. In the absence of disorder, i.e. in the limit $1/\tau_e \rightarrow 0$, $K(k)$ is given by:

\begin{equation}\label{eq23}
K(k) = \frac{\textup{Tanh}(\xi_{k}/2T_c^0)}{2 \xi_{k}},
\end{equation}
where $T_c^0$ is the critical temperature in the absence of disorder.

\section{S-wave scattering of dipolar Fermi gases in a deep 2D lattice}\label{sec3}

We focus on the 2D quadratic lattice with $N$ lattice sites. In a sufficiently deep lattice (tight-binding limit), the lowest energy band wavefunction assumes a Wannier form with an extension $\xi_0 \ll b$:

\begin{equation}\label{eq24}
\chi_\mathbf{k} (\mathbf{r}) = \frac{1}{\sqrt{N}\sqrt{\pi} \xi_0} \sum_i \textup{exp} \left[ -\frac{(\mathbf{r} - \mathbf{R}_i)^2}{2 \xi_0^2} \right] \textup{exp} \left[ i \mathbf{k R}_i \right],
\end{equation}
and the effective mass is larger compared to the mass in free space: $m_* >m$.

The s-wave scattering amplitude in 2D dipolar Fermi gases reads

\begin{equation}\label{eq25}
f(k',k) = F_0 + f_{dd} (k', k).
\end{equation}
The first term $F_0$ comes from the short-range contact interaction and can be treated as momentum independent \cite{Superfluid}. The second term comes from the long-range dipole-dipole interaction. In the case where the dipole moments $d_0$ of different components are equal, and the particle dipoles are oriented perpendicular to the plane of their translational motion, the dipole-dipole interaction potential $V_{dd} (r)$ is written as

\begin{equation}\label{eq26}
V_{dd}(r) = \frac{d_0^2}{r^3}.
\end{equation}
The long-range $1/r^3$ tail leads to the dependence of the scattering amplitude $f_{dd}(k',k)$ on the momentum, which violates the Anderson theorem and may cause critical temperature change in the presence of weak disorder.

It was shown that in the tight-binding and low momentum limits that we consider, the momentum dependent dipole-dipole part of the scattering amplitude in the lattice is the same as in free space but with replacement $m \rightarrow m_*$ \cite{Fedorov2}. Its s-wave part in the Born approximation is given

\begin{equation}\label{eq27}
\begin{split}
& f_{dd}(k', k)  = \int_{0}^{\infty} (J_0(k'r) J_0(kr) - 1) V_{dd}(r) 2 \pi r d r  \\
&= -2 \pi d_0^2
\left\{\begin{matrix}
k F(-1/2, -1/2, 1, k'^2/k^2), \:\: k'<k, \\
k' F(-1/2, -1/2, 1, k^2/k'^2), \:\: k<k'.
\end{matrix}\right.
\end{split}
\end{equation}
where $J_0$ is the Bessel function, and $F$ is the hypergeometric function. The Born approximation requires the condition $k r_{*}^l \ll 1$ with $r_{*}^l= m_* d_0^2$ being the so-called dipole-dipole distance in the lattice.

\section{Disoder-induced increase of the critical temperature}\label{sec4}
One can see that in a sufficiently deep lattice and in the low momentum limit $k_F b \ll 1$, the superfluid pairing of fermions can be considered in the same way as in free space, but we should replace $m$ with $m_*$.

\begin{figure}[h!]
\begin{center}
\center{\includegraphics[width=1\linewidth]{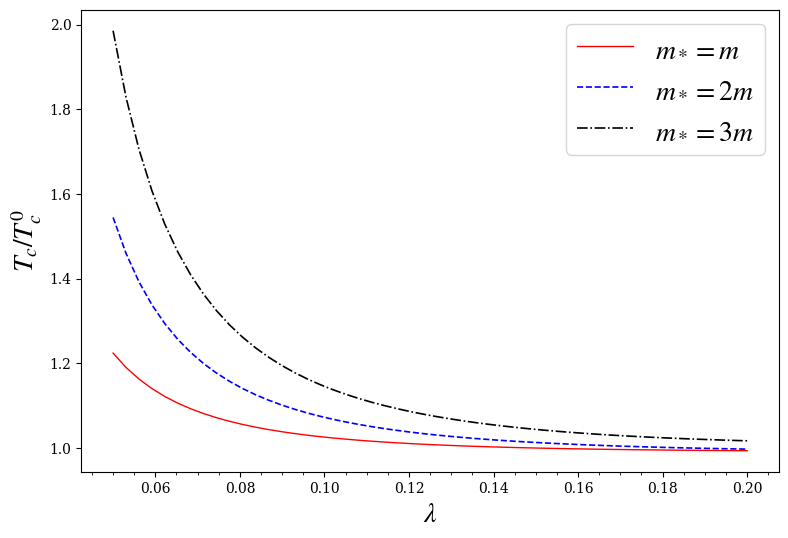}}
\end{center}
\caption{The ratio of the superfluid transition temperature in the lattice in the presence of weak disorder $T_c$ to the temperature in the absence of disorder  $T_c^0 $ versus $\lambda$. The red solid curve shows $T_c/T_c^0$ in free space ($m_* = m$), and the blue dashed (black dash-dotted) curve corresponds to $m_* = 2 m$ ($m_* = 3 m$). }
\label{results2}
\end{figure}

Figure \ref{results2} shows the numerical solution of the gap equation (\ref{eq21})
for the ratio of the  critical temperature $T_c$ in the lattice with disorder to the critical temperature 
$T_c^0$ in the lattice without disorder.
The dimensionless coupling constant  $\lambda = |f(k_F, k_F)|m/(2 \pi)$  characterizes the interaction strength. In the weakly interacting regime we have  $\lambda \ll 1$. The dipole-dipole interaction and the disorder are chosen such that $k_F$$r_{*}$ $ = 0.1$ and $k_F $$l$$ = 54$, where $r_* = m d_0^2$ and $l$ are the dipole-dipole distance and the mean free path in free space, respectively.

\begin{figure}[h!]
\begin{center}
\center{\includegraphics[width=1\linewidth]{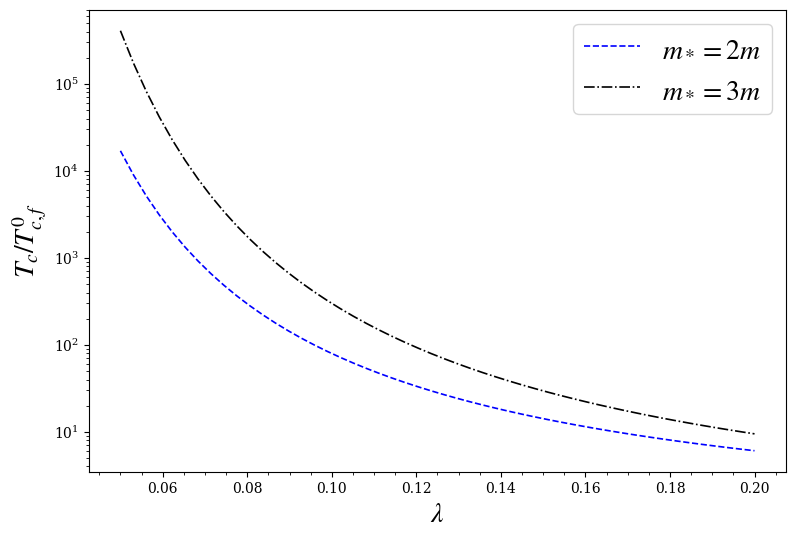}}
\end{center}
\caption{The ratio of the superfluid transition temperature in the lattice in the presence of weak disorder $T_c$ to the temperature in the absence of disorder and optical lattice $T_{c,f}^0 $ versus $\lambda$. The blue dashed (black dash-dotted) curve shows $T_c/T_{c,f}^0$ for $m_* = 2 m$ ($m_* = 3 m$). The curve for $m_* = m$ is the same as the red curve in Fig. \ref{results2}.}
\label{results1}
\end{figure}

Figure \ref{results1} shows the ratio $T_c$   to the critical temperature $T_{c,f}^0$ in free space 
without disorder.

As shown in Figs. \ref{results2} and \ref{results1}, in the presence of weak disorder the critical temperatures $T_c$ becomes larger than $T_c^0$. The increase becomes significantly more pronounced when the effective mass (depth of the lattice potential) increases. The increase of the effective mass ($m^* = \alpha m,\,  \alpha > 1$) leads to scaling of all other parameters, such as the density of states at the Fermi surface ($\nu= \frac{m}{2 \pi} \to \nu \alpha$), the mean free path ($ l = v_F \tau \to l/\alpha^2$, since $ v_F=\frac{p_F}{m} \to v_F/\alpha, \, \tau=\frac{1}{2 \pi \nu \gamma} \to \tau /\alpha$), the effective Fermi energy ($E_F= p_F v_F/2 \to E_F/\alpha$), the dipole-dipole distance ($r_*= m d_0^2 \to r_* \alpha$), and the effective dimensionless coupling constant ($\lambda= |f| \nu \to \lambda \alpha$).

 Using expressions for critical temperatures in the absence of disorder:
  $T_{c,f}^0 \sim  E_F e^{-\frac{1}{\lambda}}$,
 $T_c^0  \sim  \frac{E_F}{\alpha} e^{-\frac{1}{\lambda \alpha}}$, and
  results of the paper \cite{Superfluid} for the leading correction  to the critical temperature 
  in free space:
  $\ln \frac{T_{c,f}}{T_{c,f}^0} \sim \frac{2 r_*}{\pi^2 l}\frac{1}{\lambda^3}$,
   we can estimate
the  ratio of $T_c/T_{c,f}^0$ as
\begin{equation}
\log \frac{T_c}{T_{c,f}^0} \approx \frac{2 r_*}{\pi^2 l}\frac{1}{\lambda^3}+ \frac{1}{\lambda}\left(1-\frac{1}{\alpha}\right) - \ln \alpha.
\end{equation}


\section{Conclusions}\label{sec5}

We have investigated the influence of weak disorder on the superfluid properties of weakly interacting dipolar fermions in a 2D periodic lattice. We have found that the disorder-induced change of the superfluid transition temperature strongly depends on the lattice depth, namely, the influence of the disorder increases with increasing the lattice depth (effective mass). It is interesting to note that using both lattice and disorder one can increase the critical temperature by an order of magnitude for realistic parameters of the system ($\lambda \sim 0.2$).

\begin{acknowledgments}
This work was supported by the Russian Science Foundation Grant No. 20-42-05002.
\end{acknowledgments}

\appendix*

\section{}

The averaged Green functions are determined by the Gor'kov equations \cite{Gorkov}:

\begin{equation}\label{eq12}
(i \omega_j - \xi_k - \Sigma_G) \bar{G} + (\Delta_\mathbf{k} + \Sigma_{F}) \bar{F}^{\dagger}  =  1,
\end{equation}

\begin{equation}\label{eq13}
(i \omega_j + \xi_k - \Sigma_G) \bar{F}^{\dagger} + (\Delta_\mathbf{k} + \Sigma_{F}) \bar{G} =  0,
\end{equation}
where we omitted arguments ($\mathbf{k}$, $\omega_j$) for brevity. The self-energies due to the disorder scattering $\Sigma_G$ and $\Sigma_{F}$ are momentum independent and are given by

\begin{equation}\label{eq14}
\Sigma_G(\omega_j) = \gamma \int \frac{d^2 \mathbf{k}}{(2 \pi)^2} \bar{G}(\mathbf{k}, \omega_j),
\end{equation}

\begin{equation}\label{eq15}
\Sigma_{F}(\omega_j) = \gamma \int \frac{d^2 \mathbf{k}}{(2 \pi)^2} \bar{F}^{\dagger} (\mathbf{k}, \omega_j).
\end{equation}
At ultralow temperatures we confine ourselves to the s-wave scattering and, hence, we replace $\Delta(\mathbf{k})$ with its s-wave part $\Delta(k)$. We then solve Eqs. (\ref{eq12}) and (\ref{eq13}) in a usual manner (see, e.g., Refs. \cite{AG, our}) and find

\begin{equation}\label{eq16}
\bar{G}(\mathbf{k},\omega_j) = - \frac{i  \tilde{\omega}_j + \xi_k}{ \tilde{\omega}^2_j + \xi^2_k + \tilde{\Delta}^2_k },
\end{equation}

\begin{equation}\label{eq17}
\bar{F}(\mathbf{k},\omega_j) =  \frac{\tilde{\Delta}_k}{ \tilde{\omega}^2_j + \xi^2_k + \tilde{\Delta}^2_k },
\end{equation}
where $\tilde{\omega}_j$ and $\tilde{\Delta}_k$ are given by

\begin{equation}\label{eq18}
\tilde{\omega}_j = \omega_j \left ( 1 + \frac{1}{2 \tau_e \sqrt{\omega_j^2 + \Delta^2} } \right ),
\end{equation}

\begin{equation}\label{eq19}
\tilde{\Delta}_k= \Delta(k) + \frac{\Delta}{2 \tau_e \sqrt{\omega_j^2 + \Delta^2} },
\end{equation}
with $\Delta \equiv \Delta(k_F)$.

\end{document}